\documentclass[12pt]{article}

\usepackage{hlcite}
\usepackage{times}
\usepackage{newtxmath}
\usepackage{siunitx}
\usepackage{amsmath}
\usepackage{gensymb}
\usepackage{mathtools}
\usepackage{caption}
\usepackage{multirow}
\usepackage{hyperref}

\AtBeginDocument{
	\usepackage{booktabs}
	\usepackage{tabularx}
	\newcolumntype{C}{>{\centering\arraybackslash}X} 
}

\newcommand{\beginsupplement}{%
        \setcounter{table}{0}
        \renewcommand{\thetable}{S\arabic{table}}%
        \setcounter{figure}{0}
        \renewcommand{\thefigure}{S\arabic{figure}}%
     }

\newenvironment{hlabstract}{%
\begin{quote} \bf}
{\end{quote}}

\title{$f$-electron hybridised metallic Fermi surface in magnetic field-induced metallic YbB$_{12}$}

\author
{H. Liu,$^{1\dagger}$ A. J. Hickey,$^{1\dagger}$ M. Hartstein,$^{1}$ A. J. Davies,$^{1}$\\ A. G. Eaton,$^{1}$ T. Elvin,$^{1}$ E. Polyakov,$^{1}$ T. H. Vu,$^{1}$ V. Wichitwechkarn,$^{1}$\\ T. F\"orster,$^{2}$ J. Wosnitza,$^{2,3}$ T. P. Murphy,$^{4}$ N. Shitsevalova,$^{5}$\\ M. D. Johannes,$^{6}$ M. Ciomaga Hatnean,$^{7}$ G. Balakrishnan,$^{7}$\\ G. G. Lonzarich,$^{1}$ Suchitra E. Sebastian$^{1\ast}$\\

\\
\normalsize{$^{1}$Cavendish Laboratory, University of Cambridge,} \\
\normalsize{JJ Thomson Avenue, Cambridge, CB3 0HE, UK.}\\
\normalsize{$^{2}$ Dresden High Magnetic Field Laboratory (HLD-EMFL)} \\
\normalsize{and W\"urzburg-Dresden Cluster of Excellence ct.qmat,} \\
\normalsize{Helmholtz Zentrum Dresden Rossendorf,} \\
\normalsize{Bautzner Landstrasse 400, Dresden, 01328, Germany.} \\
\normalsize{$^{3}$ Institut f\"ur Festk\"orper- und Materialphysik,} \\
\normalsize{Technische Universit\"at Dresden, Dresden, 01062, Germany.} \\
\normalsize{$^{4}$ National High Magnetic Field Laboratory, Tallahassee, Florida, 32310, USA.} \\
\normalsize{$^{5}$ The National Academy of Sciences of Ukraine, Kiev, 03680, Ukraine.} \\
\normalsize{$^{6}$ Center for Computational Materials Science, Naval Research Laboratory,} \\
\normalsize{Washington, DC, 20375, USA.} \\
\normalsize{$^{7}$ Department of Physics, University of Warwick, Coventry, CV4 7AL, United Kingdom.} \\
\\
\normalsize{$^\ast$To whom correspondence should be addressed: suchitra@phy.cam.ac.uk.}
\\
\normalsize{$^\dagger$These authors contributed equally to this work.}
}

\date{}

\begin{document}

\baselineskip24pt

\maketitle

\clearpage

\begin{hlabstract}
  
The nature of the Fermi surface observed in the recently discovered family of unconventional insulators starting with SmB$_6$ and subsequently YbB$_{12}$ is a subject of intense inquiry. Here we shed light on this question by comparing quantum oscillations between the high magnetic field-induced metallic regime in YbB$_{12}$ and the unconventional insulating regime. In the field-induced metallic regime beyond 47~T, we find prominent quantum oscillations in the contactless resistivity characterised by multiple frequencies up to at least 3000~T and heavy effective masses up to at least 17~$m_\text{e}$, characteristic of an $f$-electron hybridised metallic Fermi surface. The growth of quantum oscillation amplitude at low temperatures in electrical transport and magnetic torque in insulating YbB$_{12}$ is closely similar to the Lifshitz-Kosevich low temperature growth of quantum oscillation amplitude in field-induced metallic YbB$_{12}$, pointing to an origin of quantum oscillations in insulating YbB$_{12}$ from in-gap neutral low energy excitations. The field-induced metallic regime of YbB$_{12}$ is characterised by more Fermi surface sheets of heavy quasiparticle effective mass that emerge in addition to the heavy Fermi surface sheets yielding multiple quantum oscillation frequencies below 1000~T observed in both insulating and metallic regimes. We thus observe a heavy multi-component Fermi surface in which $f$-electron hybridisation persists from the unconventional insulating to the field-induced metallic regime of YbB$_{12}$, which is in distinct contrast to the unhybridised conduction electron Fermi surface observed in the case of the unconventional insulator SmB$_6$. Our findings require a different theoretical model of neutral in-gap low energy excitations in which the $f$-electron hybridisation is retained in the case of the unconventional insulator YbB$_{12}$.
  
\end{hlabstract}

\section*{Introduction}

The origin of bulk quantum oscillations in bulk insulating unconventional insulators, first discovered in SmB$_6$~\cite{Tan349.287}, has been the subject of much debate~\cite{Tan349.287, Hartstein14.166, Li346.1028, Liu30.16LT01, Liu95.075426, Xiang362.65, Wang589.225, Hartstein23.101632}. Another recently discovered unconventional insulator is the Kondo insulator YbB$_{12}$~\cite{Liu30.16LT01, Xiang362.65}, in which high magnetic fields dramatically reduce the electrical resistivity, causing the metallic ground state to be realised beyond $\mu_0H\approx$~47~T~\cite{Sugiyama57.3946, Iga200.012064}. Quantum oscillation measurements in metallic YbB$_{12}$ accessed in high magnetic fields thus uniquely enable us to make a comparison between quantum oscillations in the unconventional insulating state and the field-induced metallic state.

In this paper, we experimentally compare quantum oscillations in the unconventional insulating regime and the field-induced metallic regime of YbB$_{12}$ accessed through high applied magnetic fields up to 68~T. In the field-induced metallic phase of YbB$_{12}$, we observe prominent quantum oscillations with a multiplicity of frequencies characterised by moderately heavy quasiparticle effective masses, which reflect an $f$-electron hybridised metallic Fermi surface. In order to reliably extract information from the complex quantum oscillation spectrum comprising multiple frequencies, we focus on (i) a comparison of the multiple quantum oscillation frequencies observed in both magnetic torque and electrical resistivity of the unconventional insulating regime~\cite{Liu30.16LT01} and contactless resistivity of the field-induced metallic regime, (ii) the temperature dependent quantum oscillation amplitude that can be used to distinguish between gapped and gapless Fermi surface models in the unconventional insulating regime, and (iii) consequently shed light on the nature of hybridisation in the unconventional insulating and field-induced metallic regimes.

\section*{Results}

\begin{figure}[t!]
    \vspace{-50pt}
    \begin{center}   
    \includegraphics[width=1.0\textwidth]{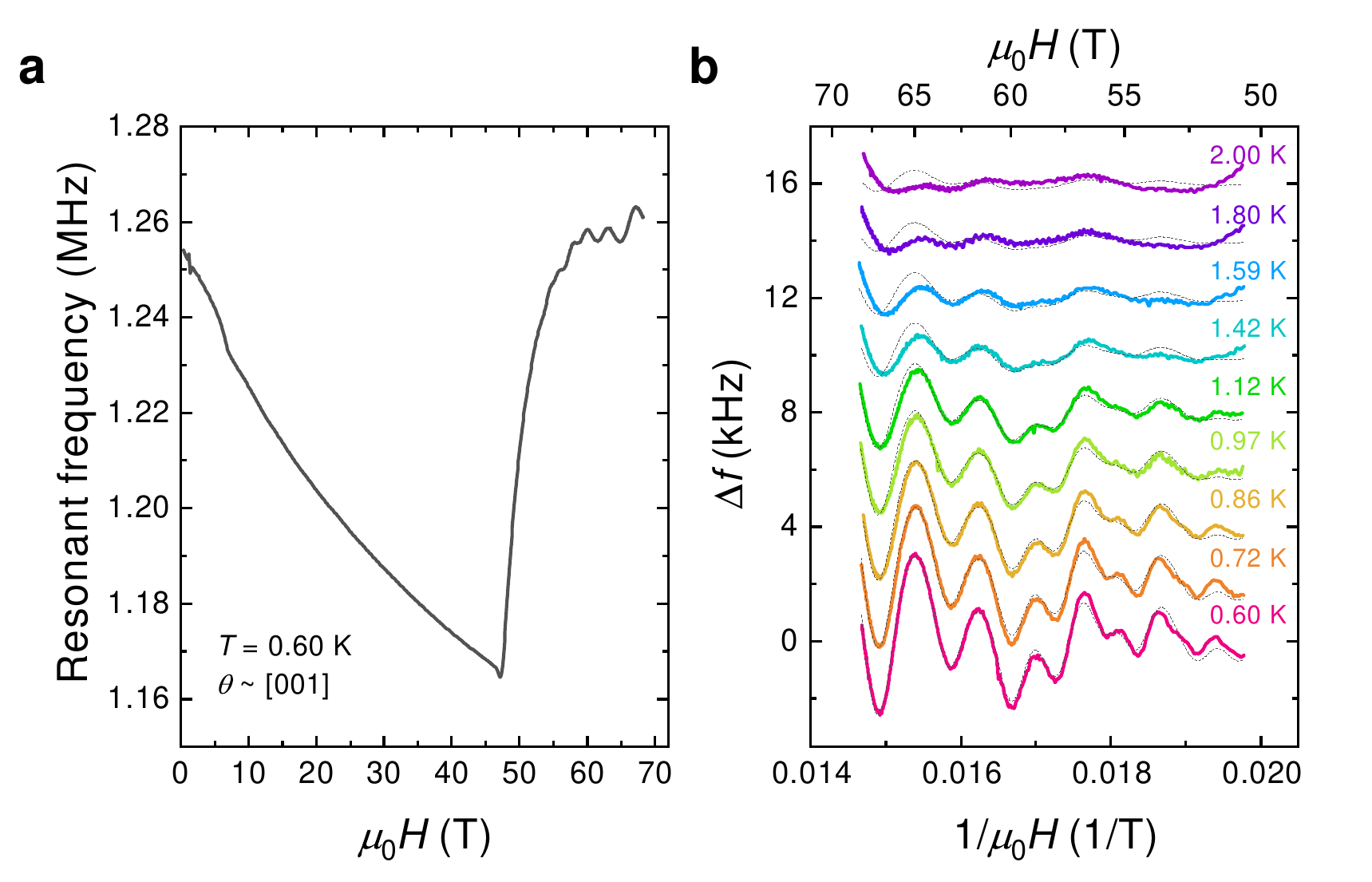}
    \end{center}
    \vspace{-20pt}
    \caption{\textbf{Quantum oscillations in the field-induced metallic phase of YbB$_{12}$.} Electrical resistivity of a single crystal of YbB$_{12}$ measured with the contactless proximity detector oscillator (PDO) technique. \textbf{(a)} PDO resonant frequency as a function of applied magnetic field up to 68~T at a temperature of 0.60~K with the field aligned close to the [001] crystallographic direction. The sharp change in resonant frequency at $\mu_0 H \approx$ 47~T indicates the onset of the insulator-metal transition~\cite{Sugiyama57.3946, Iga200.012064}. Prominent oscillations can be seen against the unsubtracted background above $\mu_0 H \approx$ 50~T. \textbf{(b)} Solid lines show quantum oscillations measured with PDO up to 68~T with a smooth monotonic background subtracted at different temperatures, where $\Delta f$ is the change in resonant frequency after background subtraction. Dashed lines show Lifshitz-Kosevich simulations of quantum oscillations using multiple frequency components as listed in Table~\ref{table:masses}.}
    \label{pic:QO}
\end{figure}

\begin{figure}[t!]
    \vspace{-50pt}
    \begin{center}   
    \includegraphics[width=1.0\textwidth]{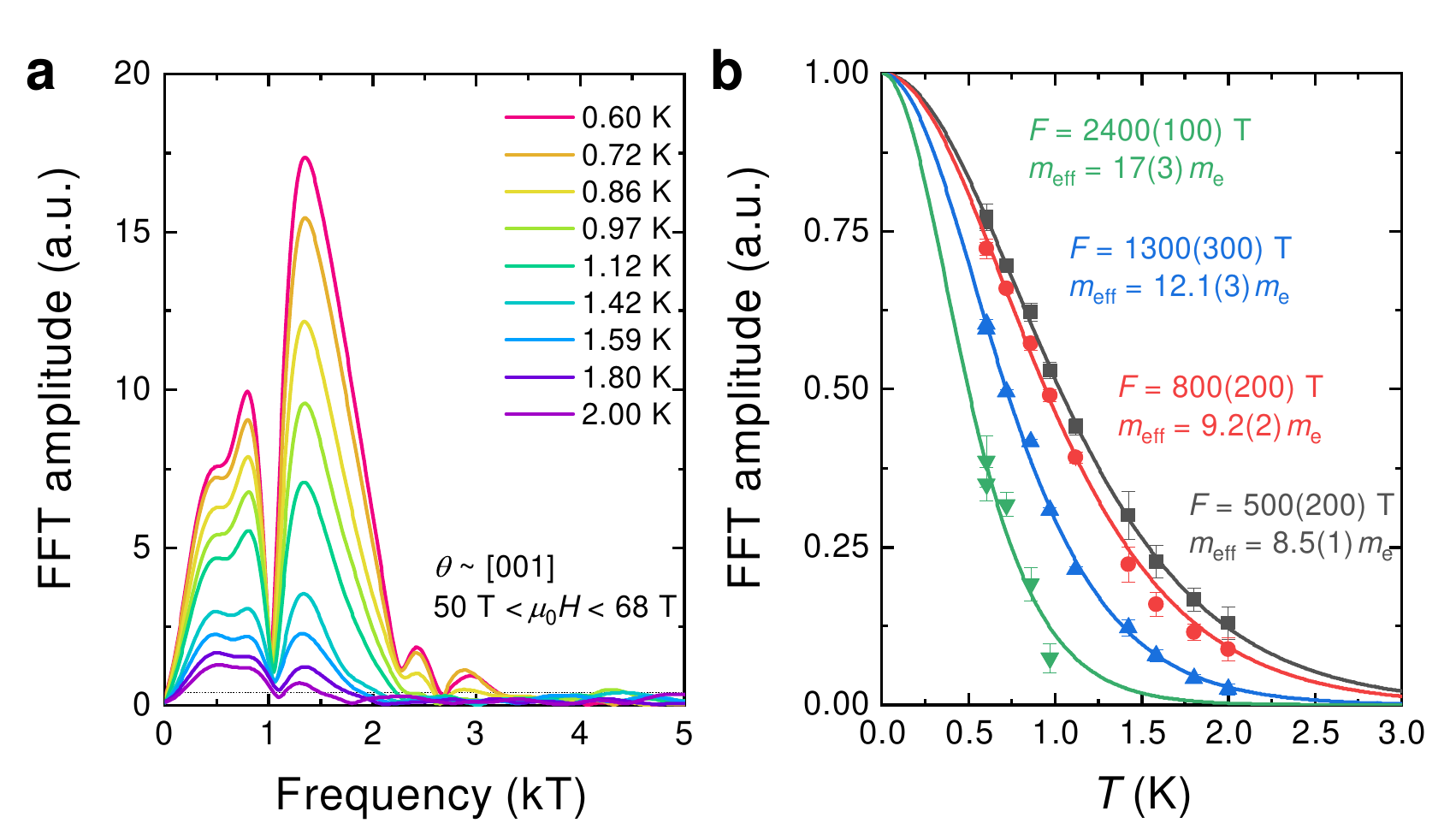}
    \end{center}
    \vspace{-20pt}
    \caption{\textbf{Rich spectrum of multiple quantum oscillation frequencies in the field-induced metallic phase of YbB$_{12}$.} \textbf{(a)} Fourier transforms of the subtracted quantum oscillations shown in Fig.~\ref{pic:QO}b, for a field window of $50~\text{T} < \mu_0 H < 68~\text{T}$ at temperatures between 0.6 -- 2.0~K, where the applied field was aligned close to the [001] crystallographic direction. The horizontal dashed line indicates the FFT noise floor. Multiple distinct quantum oscillation frequency peaks between 500(200)~T and at least 3000(200)~T are visible. \textbf{(b)} Quantum oscillation amplitude obtained from the Fourier transform peak height shown in (a) as a function of temperature for multiple representative frequencies. Lifshitz-Kosevich temperature dependence fits, as shown by the solid lines, yield quasiparticle effective masses $m^* / m_\text{e}$  between 8.5(1) -- 17(3) for the various frequencies. A summary of the multiple observed frequencies and their corresponding effective masses is shown in Table~\ref{table:masses}.}
    \label{pic:FFT}
\end{figure}

Figure~\ref{pic:QO}a shows quantum oscillations in the contactless electrical resistivity of a single crystal of YbB$_{12}$ measured using the proximity detector oscillatory (PDO) technique, at high magnetic fields above the insulator-metal transition at $\mu_0H\approx$ 47~T~\cite{Iga200.012064}. Prominent quantum oscillations are visible in the measured contactless electrical resistivity before background subtraction. Figure~\ref{pic:QO}b shows quantum oscillations after smooth, monotonic backgrounds have been subtracted from the contactless electrical resistivity (measured by the resonant frequency) above 50~T at various temperatures, where the quantum oscillation periodicity in inverse magnetic field can be seen. Multiple frequency peaks between 500(200)~T -- 3000(200)~T are revealed by Fast Fourier Transforms (FFT) of the background-subtracted quantum oscillations, as shown in Fig.~\ref{pic:FFT}a. Plotting the quantum oscillation amplitude as a function of temperature down to 0.6~K yields a Lifshitz-Kosevich (LK) temperature dependence with cyclotron effective masses $m^* / m_\text{e}$ between 8.5(1) -- 17(3), as shown in Fig.~\ref{pic:FFT}b.

\section*{Discussion}

\begin{figure}[t!]
    \vspace{-20pt}
    \begin{center}
    \includegraphics[width=1\textwidth]{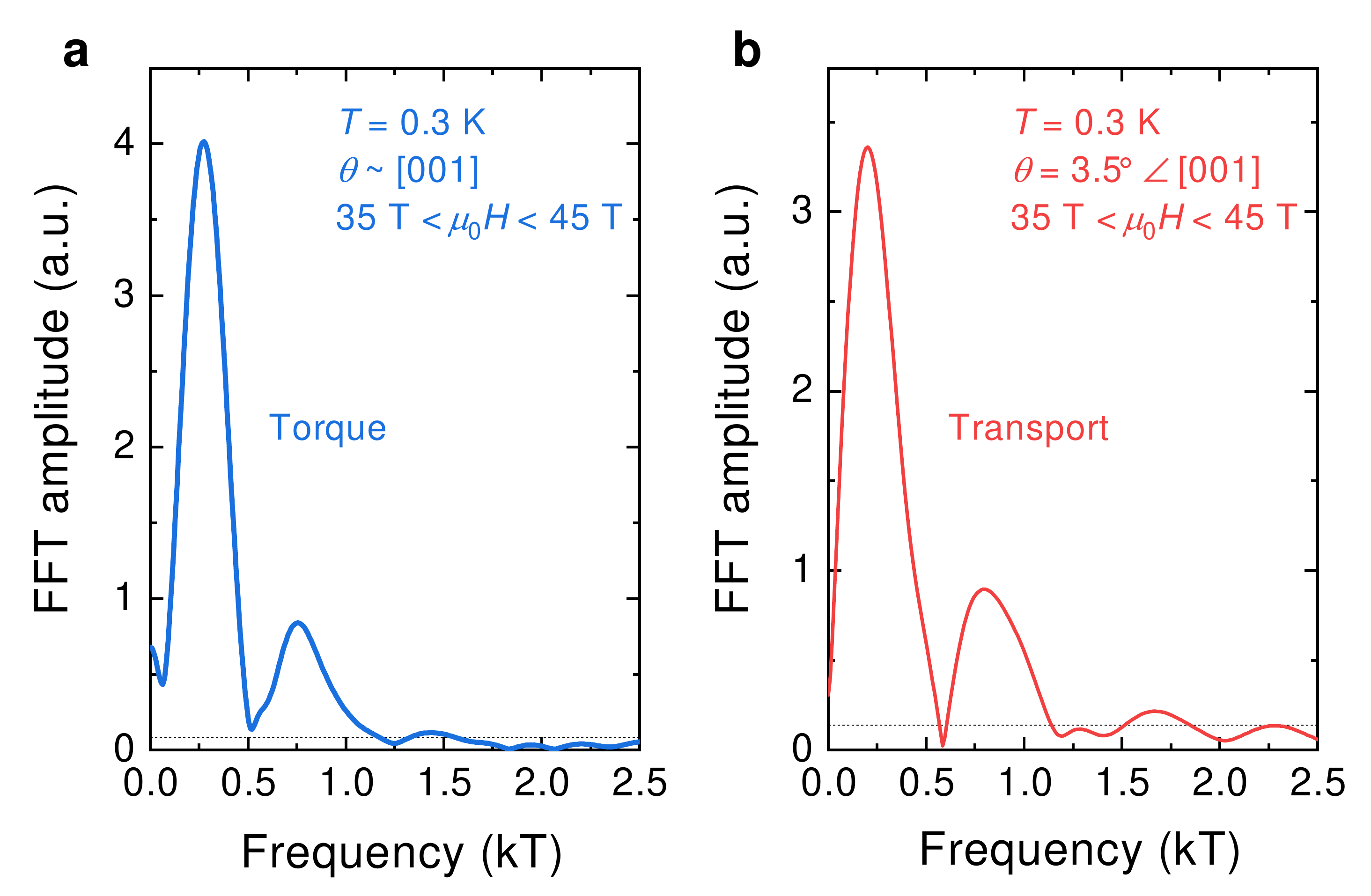}
    \end{center}
    \vspace{-20pt}
    \caption{\textbf{FFT of quantum oscillations in insulating YbB$_{12}$.} FFT of quantum oscillations measured on insulating YbB$_{12}$ at a temperature of 0.3~K and field range of 35~T~$< \mu_0 H < $~45~T in \textbf{(a)} magnetic torque, with the applied field aligned close to the [001] crystallographic direction, and \textbf{(b)} electrical resistivity, with the applied field aligned 3.5$^\circ$ away from the [001] crystallographic direction in the [001]-[111]-[110] rotation plane. The horizontal dashed lines indicate the FFT noise floors. Similar frequencies can be discerned in the two measured physical quantities. A summary of the multiple observed frequencies and their respective quasiparticle effective masses is shown in Table~\ref{table:masses}. An FFT decomposition involving LK simulations identifies a frequency $\approx$~450~T in electrical resistivity that is visible as a shoulder of the main peak.}
    \label{YbB12_insulating_FFTs}
\end{figure}

Figure~\ref{YbB12_insulating_FFTs} shows multiple quantum oscillation frequencies in the insulating phase of YbB$_{12}$ measured through capacitive magnetic torque and contacted electrical transport~\cite{Liu30.16LT01}. Fig.~\ref{pic:FFT}a shows the quantum oscillation frequency spectrum in the field-induced metallic phase, comprising multiple frequencies extending up to at least 3000(200)~T. We note that even higher frequencies may exist, especially in view of the high value of linear specific heat $\gamma \approx 67$~mJ~mol$^{-1}$~K$^{-2}$ measured in the field-induced metallic regime of YbB$_{12}$~\cite{Terashima120.257206}. Multiple comparable quantum oscillation frequencies between approximately 300 -- 800~T are measured in both the metallic and insulating phases (Table~\ref{table:masses}); while multiple frequencies were previously reported in the insulating phase of YbB$_{12}$ measured through capacitive magnetic torque~\cite{Liu30.16LT01}, these were missed in other reports of a single quantum oscillation frequency in the insulating phase of YbB$_{12}$~\cite{Xiang362.65}. Such a quantum oscillation spectrum comprising multiple frequencies is expected from numerical Fermi surface simulations of metallic YbB$_{12}$ involving hybridised $f$-electrons~\cite{Liu30.16LT01}. In these theoretical simulations of the Fermi surface, multiple Fermi surface pockets located away from the centre of the Brillouin zone would be expected to yield a series of frequency branches; multiple frequencies would further be expected from a multiplicity of electron and hole pockets. The multiple quantum oscillation frequencies experimentally observed in both the magnetic torque and electrical resistivity of the unconventional insulating phase and the contactless resistivity of the field-induced metallic phase yield a complex quantum oscillation spectrum that unfortunately cannot be treated by analysis methods such as Landau level indexing in inverse magnetic field, complete mapping of the Fermi surface geometry, and other traditional Fermi surface treatments. In this work, therefore, we focus instead on a robust treatment involving comparison between the multiple quantum oscillation frequencies and the temperature-dependence of the quantum oscillation amplitudes observed in both the unconventional insulating~\cite{Liu30.16LT01} and field-induced metallic regimes.

Broad classes of models that have been proposed to explain bulk quantum oscillations in unconventional insulators include categories of gapped models, and models characterised by in-gap low energy excitations~\cite{Miyake186.115, Knolle115.146401, Riseborough96.195122, Peters100.085124, Zhang116.046404, Lee103.L041101, Bulaevskii78.024402, Motrunich73.155115, Baskaran1507.03477, Erten119.057603, Varma102.155145, Chowdhury9.1766}. An analysis of the temperature-dependence of the quantum oscillation amplitude provides us with vital information to distinguish between classes of gapped and gapless models to describe quantum oscillations in the unconventional insulating phase.

\begin{figure}[t!]
    \vspace{-20pt}
    \begin{center}
    \includegraphics[width=1\textwidth]{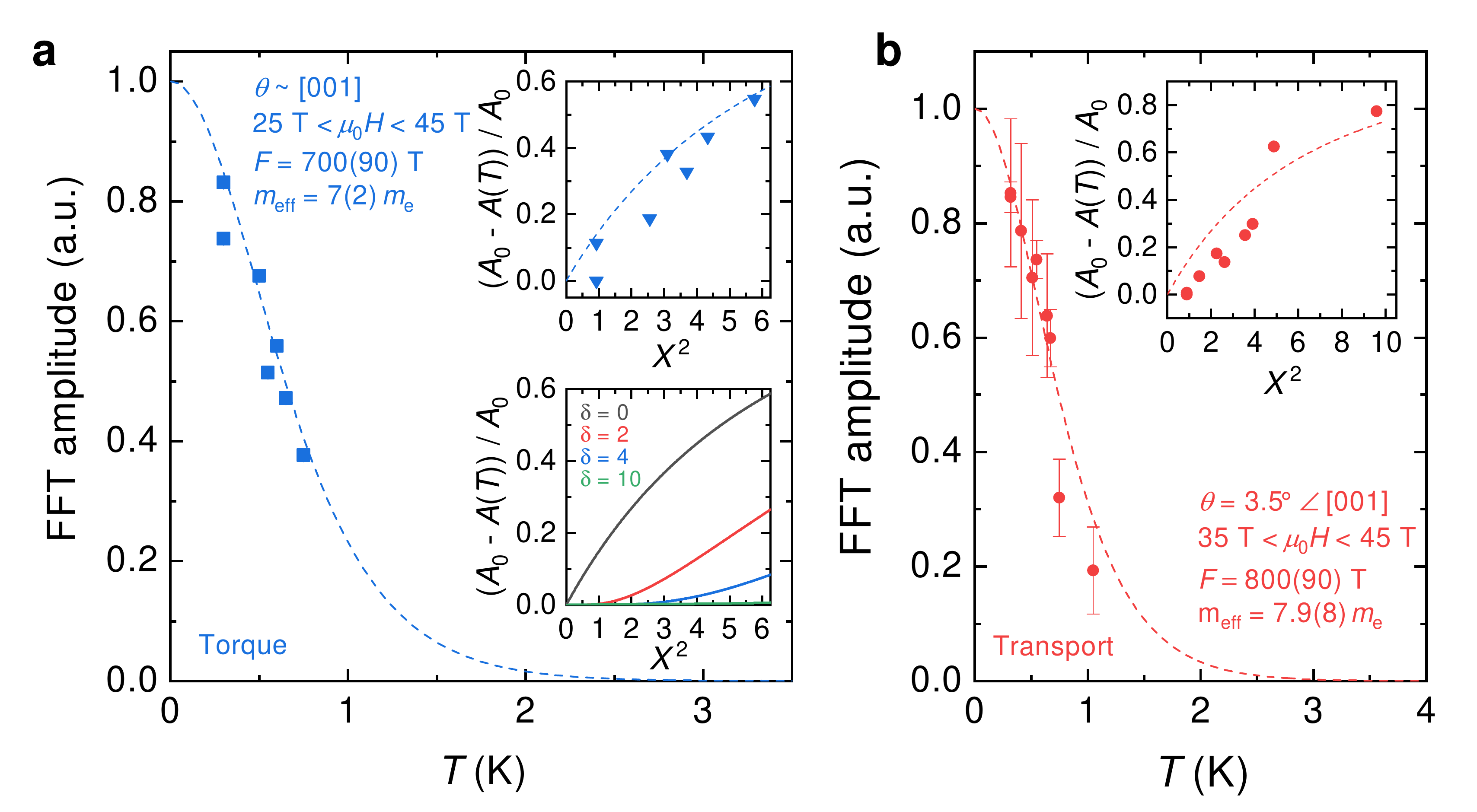}
    \end{center}
    \vspace{-20pt}
    \caption{\textbf{Gapless low energy excitations yield low temperature Lifshitz-Kosevich growth of quantum oscillation amplitude in the insulating phase of YbB$_{12}$.} \textbf{(a)} Amplitude of the 700~T frequency quantum oscillations measured using cantilever torque magnetometry as a function of temperature, with the applied field aligned close to the [001] crystallographic direction. Measured quantum oscillation amplitude follows the Lifshitz-Kosevich (LK) form (dashed line) down to lowest measured temperatures. (Lower inset) Low temperature model expansion of quantum oscillation amplitude from refs.~\cite{Miyake186.115,Knolle115.146401} shows non-LK activated behaviour for various finite gap sizes ($\delta\approx12$ for YbB$_{12}$ at 40~T~\cite{Sugiyama57.3946}), in contrast to LK exponential growth expected for gapless low energy excitations ($\delta=0$). Model calculations are shown in Supplementary Information. $A(T)$ is the quantum oscillation amplitude at temperature $T$, $A_0$ is the amplitude at the lowest measured temperature, $X = 2 \pi^2 k_\text{B} T / \hbar\omega_\text{c}$ is the temperature damping coefficient in the LK formula~\cite{Shoenberg1984}, $\delta = {2 \pi \Delta} / {\hbar \omega_\text{c}}$ where $\Delta$ is the isotropic gap size and $\omega_{\rm c}$ is the cyclotron frequency. (Upper inset) Growth of magnetic torque quantum oscillation amplitude at the lowest measured temperatures; experimental data (solid triangles) exhibits good agreement with gapless model simulation (dashed lines). \textbf{(b)} Amplitude of the 800~T frequency quantum oscillations measured using four-point contacted electrical transport as a function of temperature, with the applied field aligned 3.5$^\circ$ away from the [001] crystallographic direction in the [001]-[111]-[110] rotation plane. Measured quantum oscillation amplitude follows LK form (dashed line) down to lowest measured temperatures. (Inset) Growth of electrical transport quantum oscillation amplitude at lowest temperatures; experimental data (solid circles) exhibits good agreement with gapless model simulation (dashed line).}
    \label{YbB12_lowTExp}
\end{figure}

At the simplest level of weakly interacting gapped systems, the system is characterised by a single particle gap. Models in this category have for example been proposed for BCS superconductors~\cite{Miyake186.115} and for weakly interacting insulators~\cite{Knolle115.146401, Riseborough96.195122}. Examples of such behaviour are experimentally observed in unconventional superconductors~\cite{Settai13.L627,Hedo77.975,Inada68.3643,Ohkuni79.1045}. For this category of gapped models of quantum oscillations in weakly interacting insulators, the quantum oscillation amplitude exhibits a non-LK flattening or decrease at low temperatures~\cite{Miyake186.115,Knolle115.146401} (Supplementary Information, Figure~\ref{YbB12_lowTExp}a lower inset). Other models of weakly interacting gapped systems invoke quantum oscillations arising from modulation of the gap resulting from an inverted band structure~\cite{Peters100.085124, Zhang116.046404, Lee103.L041101}.

This picture is modified in the case of strongly correlated insulators. These insulators, driven by strong interactions, are expected to be characterised by an in-gap density of states, as predicted by various theoretical models. For instance, models of single-band Mott insulators~\cite{Bulaevskii78.024402, Motrunich73.155115} involve low energy excitations of chiefly spin character. Models of Majorana fermions proposed for Kondo insulators include those in refs.~\cite{Baskaran1507.03477, Erten119.057603, Varma102.155145}. In these models, low energy excitations involve Majorana fermion bands, that can be a linear equal combination of a canonical particle and anti-particle operators, crossing the chemical potential. Another model has been proposed for quantum oscillations from composite fermionic excitons in Kondo insulators~\cite{Chowdhury9.1766}. In this case, mixed-valence insulators are proposed to host a fractionalised neutral Fermi sea, which develops an emergent magnetic field in the presence of a physical magnetic field. In these various models, the Fermi surface may be expected to correspond to the unhybridised conduction electron band. Further, the quantum oscillation amplitude in these gapless models is expected to increase at low temperatures, for instance obeying an LK form in the case of low energy excitations characterised by Fermi-Dirac statistics (Supplementary Information, Figure~\ref{YbB12_lowTExp}a lower inset).

Figure~\ref{pic:FFT} shows the quantum oscillation amplitude as a function of temperature in field-induced metallic YbB$_{12}$, growing in accordance to the LK form down to the lowest measured temperatures, as expected for a metal characterised by Fermi-Dirac statistics. We obtain the cyclotron effective mass for multiple quantum oscillation frequencies in the field-induced metallic phase of YbB$_{12}$ from an LK fit to the quantum oscillation amplitude as a function of temperature (Fig.~\ref{pic:FFT}b). Table~\ref{table:masses} shows a range of moderately high effective masses $m^* / m_\text{e}$ up to at least 17(3) observed for multiple quantum oscillation frequencies up to at least 3000(200)~T. The heavy effective masses observed in the field-induced metallic phase indicate its correspondence to an $f$-electron hybridised metallic Fermi surface~\cite{Terashima120.257206}.

The presence of neutral low-energy excitations in the gap would be expected to yield an increase in quantum oscillation amplitude at low temperatures in strongly correlated models, which distinguishes them from gapped models of quantum oscillations in weakly interacting insulators in which the quantum oscillation amplitude is expected to exhibit non-LK flattening or decrease at low temperatures~\cite{Miyake186.115, Knolle115.146401} (Supplementary Information, Figure~\ref{YbB12_lowTExp}a lower inset). Figure~\ref{YbB12_lowTExp} shows the temperature dependence of quantum oscillation amplitude for multiple representative frequencies in magnetic torque and electrical transport measured in the insulating phase of YbB$_{12}$~\cite{Liu30.16LT01}. Similar to our observation in the metallic phase, the quantum oscillation amplitude of both magnetic torque and electrical resistivity in the insulating phase grows in accordance with the LK form down to the lowest measured temperatures, below the gap temperature beneath which gapped models of quantum oscillations predict a non-LK flattening or decrease in amplitude~\cite{Miyake186.115,Knolle115.146401}. LK fits to the quantum oscillation amplitude as a function of temperature of quantum oscillation frequencies between 300~T and 800~T observed in the insulating phase yield moderately heavy effective masses $m^* / m_\text{e}$ between approximately 4.5 -- 9, which are similar to the effective masses observed in the field-induced metallic phase for a similar range of quantum oscillation frequencies (Table~\ref{table:masses}). The growth in quantum oscillation amplitude down to the lowest measured temperatures is clearly evidenced in the two upper insets in Fig.~\ref{YbB12_lowTExp}, which highlight low temperature growth of the torque and transport quantum oscillation amplitude measured in the insulating phase. This striking observation of a steep increase in quantum oscillation amplitude down to the lowest temperature is in clear contrast to the non-LK flattening or decrease expected for gapped Fermi surface models, a simulation of which is shown in the lower inset of Fig.~\ref{YbB12_lowTExp} for various gap values, exhibiting non-LK finite activation behaviour for a finite gap. We are thus able to identify quantum oscillation signatures in the unconventional insulator YbB$_{12}$ that reveal an origin from in-gap neutral low-energy excitations, as yielded by correlated insulator models. 

\begin{table}[t!]
	\begin{tabularx}{\textwidth}{CCCCCC}
	\toprule
	\multicolumn{4}{c}{Insulating phase} & \multicolumn{2}{c}{Metallic phase} \\
	\cmidrule(lr){1-4} \cmidrule(lr){5-6}
	Frequency (T) & Mass ($m_\text{e}$) & Frequency (T) & Mass ($m_\text{e}$) & Frequency (T) & Mass ($m_\text{e}$)\\
	\cmidrule(lr){1-2} \cmidrule(lr){3-4} \cmidrule(lr){5-6}
	\multicolumn{2}{c}{dHvA ($\theta \sim [001]$)} & \multicolumn{2}{c}{SdH ($\theta = 3.5^\circ \angle$ [001])} & \multicolumn{2}{c}{PDO  ($\theta \sim [001]$)} \\
	\cmidrule(lr){1-2} \cmidrule(lr){3-4} \cmidrule(lr){5-6}
	{} &	{} &	150(90) &	3.2(2) &	{} &	{}	\\
	300(70) &	4.5(5) &	450(80) &	6.1(6) &	500(200) &	8.5(1)	\\
	700(90) &	7(2) &	800(90) &	7.9(8) &	800(200) &	9.2(2) 	\\
	{} &		{} &	{} &	{} & 	1300(200) &	12.1(3) 	\\
	{} &		{} &	{} &		{} &		1700(200) & 	16(5) 	\\
	{} &		{} &	{} &		{} &		2300(200) & 	17(3) 	\\
	{} & 		{} &	{} &		{} &		3000(200) & 	14(3) 	\\
	\bottomrule
	\end{tabularx}
    \caption{\textbf{Observed multiple quantum oscillation frequencies and effective masses in the insulating and metallic phases of YbB$_{12}$.} Multiple quantum oscillation frequencies and cyclotron effective masses measured with capacitive torque magnetisation (de Haas-van Alphen (dHvA) oscillations) and four-point contacted resistivity (Shubnikov-de Haas (SdH) oscillations) in the insulating phase of YbB$_{12}$, and with proximity detector oscillator (PDO) contactless electrical transport in the magnetic field-induced metallic phase of YbB$_{12}$. The applied magnetic field was aligned close to the [001] crystallographic direction for dHvA and PDO measurements, and was aligned 3.5$^\circ$ from the [001] crystallographic direction in the [001]-[111]-[110] rotational plane for the SdH measurements. The FFT field range was $50~\text{T} < \mu_0 H < 68~\text{T}$ for PDO, $35~\text{T} < \mu_0 H < 45~\text{T}$ for the 700~T frequency in dHvA and 800~T frequency in SdH, and $28~\text{T} < \mu_0 H < 45~\text{T}$ for other frequencies in dHvA and SdH.}
    \label{table:masses} 
\end{table}

Our comparison of measured quantum oscillations between the unconventional bulk insulating regime~\cite{Liu30.16LT01} and field-induced metallic regime of YbB$_{12}$ shows that an application of magnetic fields yields a spectrum of multiple quantum oscillation frequencies that appear prominently in magnetic field-induced metallic YbB$_{12}$, encompassing similar frequencies below $1000$~T observed in insulating YbB$_{12}$, but extending to higher frequencies up to at least 3000(200)~T (Table~\ref{table:masses}). The comparable quantum oscillation frequency range observed in both metallic and insulating regimes is characterised by similar moderately heavy effective masses in both regimes, while higher frequencies in the field-induced metallic phase are characterised by heavy effective masses $m^\ast/m_{\rm e}$ up to at least 17(3). This appearance of multiple additional heavy Fermi surface sheets in the magnetic field-induced metallic regime of YbB$_{12}$ would explain the steep increase in the linear specific heat at the field-induced insulator metal transition reported in~\cite{Terashima120.257206}.

\begin{figure}[t!]
    \vspace{-20pt}
    \begin{center}
    \includegraphics[width=1\textwidth]{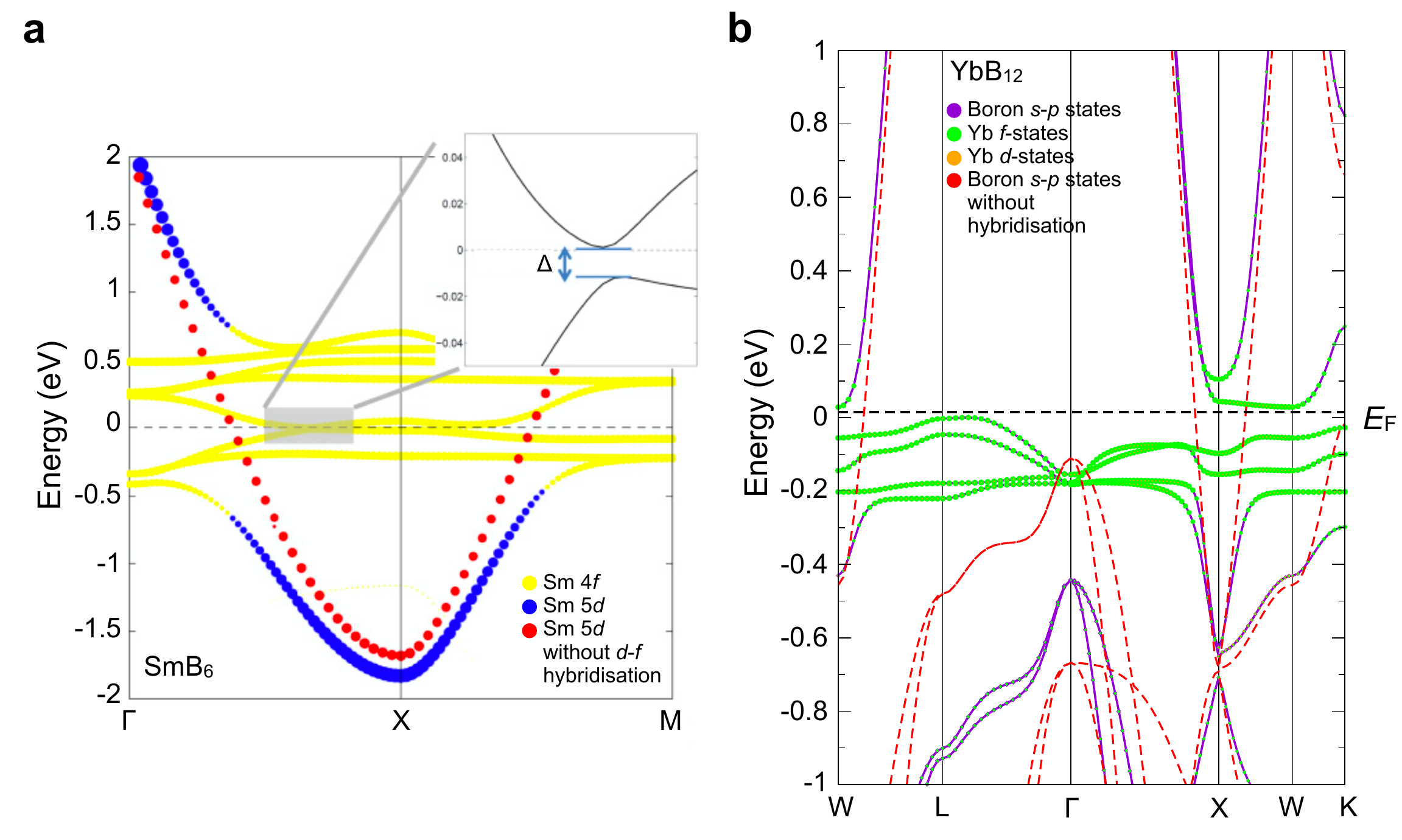}
    \end{center}
    \vspace{-10pt}
    \caption{\textbf{Contrasting band structure of SmB$_6$ and YbB$_{12}$.} \textbf{(a)} Band structure of SmB$_6$ from GGA calculations in ref.~\cite{Chen91.155151}, zoomed in view near the Fermi energy $E_\text{F}$ (full energy range shown in ref.~\cite{Chen91.155151}). Sizes of yellow and blue dots denote weights of Sm-4\textit{f} and Sm-5\textit{d} in various bands. Red dots denote metallic Sm-5\textit{d} orbitals without hybridisation with Sm-4\textit{f} orbitals. \textbf{(b)} Calculated band structure of YbB$_{12}$ shown with an expanded view around the Fermi energy $E_\text{F}$ (full energy range shown in ref.~\cite{Liu30.16LT01}). Size of the circles are proportional to the weight at each \textit{k}-point, green circles are Yb \textit{f}-states, orange circles are Yb \textit{d}-states, and violet circles are boron \textit{s-p}-states. Red dots denote two partially filled unhybridised boron \textit{s}-\textit{p} conduction electron orbitals without hybridisation with Yb-\textit{f} orbitals. In both cases, the Fermi surface yielded by the unhybridised band is not simply connected, leading to a large number of expected frequencies~\cite{Tan349.287,Liu30.16LT01}.}
    \label{bs}
\end{figure}

Our observation of a heavy Fermi surface with multiple quantum oscillation frequencies in the unconventional insulating and high field-induced metallic regimes of YbB$_{12}$ points to a multi-component Fermi surface characterised by $f$-electron hybridisation that persists from the unconventional insulating regime to the high field metallic regime. We note a crucial distinction between the band structure of unconventional insulators SmB$_6$~\cite{Tan349.287} and YbB$_{12}$~\cite{Liu30.16LT01}. While in the case of SmB$_6$, a single half-filled unhybridised conduction $d$-electron band crosses the Fermi energy and hybridises with the $f$-electron band to yield the Kondo gap (Figure~\ref{bs}a), the situation is different in YbB$_{12}$. In the case of YbB$_{12}$, two partially filled unhybridised $s$-$p$ conduction electron bands that are cumulatively half-filled cross the Fermi energy with electron-like character, and are gapped by hybridisation with the $f$-electron band (Figure~\ref{bs}b). We find this difference leads to a distinct contrast between the case of the unconventional insulator YbB$_{12}$, where heavy Fermi surface sheets are characterised by $f$-electron hybridisation, and the case of SmB$_6$, in which the observed light Fermi surface sheets correspond to an unhybridised conduction electron band~\cite{Tan349.287}. Our findings in YbB$_{12}$ are a challenge to correlated models of in-gap states that are expected to yield a Fermi surface corresponding to an unhybridised conduction electron band. An alternative possibility is suggested by the close proximity of the underlying bandstructure to a semimetallic bandstructure comprising heavy $f$-electron hybridised electron and hole pockets (Fig.~\ref{bs}). For weak correlations between electrons and holes, metallic electrical conduction would be expected. In contrast, for strong correlations, the electrons and holes may be expected to combine, such that they cannot be readily decoupled, thus impeding longitudinal electrical conduction. Despite the electrically insulating behaviour in such a strongly correlated case where electrons and holes are coupled, the Lorentz force can still drive orbital currents, which can yield quantum oscillations corresponding to a heavy $f$-electron hybridised semimetallic Fermi surface of the kind observed.

\clearpage

\section*{Methods}

\subsection*{Sample preparation}

Source polycrystalline YbB$_{12}$ powder was synthesised using borothermal reduction of 99.998\% mass purity Yb$_2$O$_3$ powder and 99.9\% mass purity amorphous B at 1700$^\circ$C under vacuum~\cite{Werheit23.065403}. The synthesized powder was isostatically pressed into a cylindrical rod and sintered at 1600$^\circ$C in Ar gas flow for several hours. Single crystals of YbB$_{12}$ were grown by the traveling solvent floating zone technique under conditions similar to those in ref.~\cite{Iga177.337} using a four-mirror Xe arc lamp (3~kW) optical image furnace from Crystal Systems Incorporated, Japan. The growths were performed in a reducing atmosphere of Ar with 3\% H$_2$ at a rate of 18~mm~hr$^{-1}$ with the feed and seed rods counter-rotating at 20--30~rpm. Samples for all measurement techniques were cut to size using a wire saw and electropolished to remove heat damage and surface strain.

\subsection*{Proximity detector oscillator}

Contactless electrical transport measurements using the proximity detector oscillator (PDO) technique~\cite{Altarawneh80.066104} were performed using a long-pulse magnet capable of generating up to 68~T at the Hochfeld Magnetlabor Dresden (HLD) in Dresden, Germany. The capacitor bank-driven magnet has a pulse duration of 150~ms, and is fitted with a custom made $^3$He system with a base temperature of $\approx$~600~mK. The PDO circuit was made in accordance to ref.~\cite{Altarawneh80.066104}, using a hand-wound sensing coil with 10~turns. The raw frequency output from the PDO circuit was $\sim$~20~MHz, which was passed through a processing circuit before being recorded at $\sim$~1~MHz using a National Instruments PXI system recording at 15~MHz.

\subsection*{Capacitive torque magnetometry}

Torque magnetometry measurements were performed in DC magnetic fields at the National High Magnetic Field Laboratory in Tallahassee, Florida, USA. The 45~T hybrid magnet was operated with a $^3$He system capable of reaching temperatures as low as 300~mK. 

Cantilevers were cut from 20~$\muup$m or 50~$\muup$m thick pieces of BeCu into flexible T-shaped pieces. Samples of dimensions approximately 1~$\times$~1~$\times$~0.5~mm$^3$ were secured on the wide end of the cantilever using epoxy, which was thermally matched to the sample to minimise strain. The narrow end of the cantilever was secured down such that the wide end of the cantilever hovers above a Cu baseplate, forming the two plates of a capacitor. The change in capacitance between the two plates was measured using a General Radio analogue capacitance bridge with a lock-in amplifier. 

\subsection*{Density Functional Theory Calculations}

Density functional theory bandstructures  were calculated with the Wien2k augmented plane wave plus local orbital (APW+lo) code~\cite{Blaha2001}. The modified Becke-Johnson (mBJ) potential was used, which is a semi-local approximation to the exact exchange plus a screening term~\cite{Tran102.226401} and which improves the band gap in many semiconductor materials. Application of mBJ resulted in a non-magnetic ground state with an indirect band gap of 21~meV and a direct gap of 80~meV, whereas the standard Perdew Burke Ernzerhof (PBE) potential produced a semimetal with overlapping valence and conduction bands. Spin-orbit coupling was included via the second variational method and resulted in a strong reordering of the bands. Self-consistent calculations were converged using a k-mesh of $15\times15\times15$ followed by a non-self-consistent calculation with a $30\times30\times30$ mesh for calculation of Fermi surfaces. The bandstructure for boron $s$-$p$ states without hybridisation was calculated by shifting the $f$-bands out of the energy range of hybridisation using DFT+U~\cite{Anisimov44.943}.

\clearpage

\bibliographystyle{naturemag}
\bibliography{YbB12_Feb_19.bib}

\section*{Acknowledgements}

H.L, A.J.H., M.H., A.J.D., A.G.E., and S.E.S. acknowledge support from the Royal Society, the Leverhulme Trust through the award of a Philip Leverhulme Prize, the Winton Programme for the Physics of Sustainability, the Taiwanese Ministry of Education, EPSRC UK (studentship and grant numbers EP/M506485/1, EP/P024947/1, EP/1805236, EP/2124504), the Royal Society of Chemistry (researcher mobility grant M19-1108), and the European Research Council under the European Unions Seventh Framework Programme (Grant Agreement numbers 337425 and 772891). M.D.J. acknowledges support for this project by the Office of Naval Research (ONR) through the Naval Research Laboratory’s Basic Research Program. M.C.H. and G.B. acknowledge financial support from EPSRC, UK through Grant EP/T005963/1. We thank the team at the National Academy of Sciences of Ukraine, Kiev for assistance in the preparation of polycrystalline YbB$_{12}$. A portion of magnetic measurements were carried out using the Advanced Materials Characterisation Suite in the University of Cambridge, funded by EPSRC Strategic Equipment Grant EP/M000524/1. 

We acknowledge support from the Deutsche Forschungsgemeinschaft (DFG) through the W\"urzburg-Dresden Cluster of Excellence on Complexity and Topology in Quantum Matter-ct.qmat (EXC 2147, Project No. 390858490) as well as the support of the HLD at HZDR, a member of the European Magnetic Field Laboratory (EMFL). A portion of this work was performed at the National High Magnetic Field Laboratory, which is supported by National Science Foundation Cooperative Agreement No. DMR-1157490, the State of Florida and the DOE.

\clearpage

\beginsupplement

\section*{Supplementary Information}

\subsection*{Non LK quantum oscillation amplitude temperature-dependence for gapped models compared with LK temperature-dependence for gapless models}

\subsubsection*{Model simulations}

To distinguish between gapless and gapped models of quantum oscillations in the unconventional insulating phase, here we simulate the quantum oscillation amplitude for various gap sizes. We use the formulation of ref.~\cite{Miyake186.115,Knolle115.146401}, the ratio of the first harmonic between the gapped state and the normal state is:
\begin{equation}
\frac{M_\text{g}}{M_\text{n}} = \frac{\sinh(X)}{X} \int_0^\infty \cos\left( \frac{X \mu}{\pi} \right) \partial_\mu \left( \frac{\mu}{\sqrt{\mu^2 + (\Delta/T)^2}} \tanh\left( \frac{\sqrt{\mu^2 + (\Delta/T)^2}}{2}\right) \right) \text{d}\mu,
\end{equation}
where $\mu$ is the chemical potential,  $\Delta$ is the isotropic gap size, and $X$ is the temperature damping coefficient given by $X = 2 \pi^2 k_\text{B} T m^* / e \hbar B_0$. Here, $k_\text{B}$ is Boltzmann's constant, $T$ is temperature, $m^*$ is the quasiparticle effective mass, $e$ is the electron charge, $\hbar$ is the reduced Planck constant, and $B_0 = \mu_0 H$ is the applied magnetic field~\cite{Shoenberg1984}. \\[-6pt]

\noindent If we set $T = X \omega_c / (2 \pi^2)$ and $\Delta / T = 2 \pi^2 \Delta / \omega_\text{c} X = \pi \delta / X$, we find:
\begin{equation}
\delta = \frac{2 \pi \Delta}{\hbar \omega_\text{c}},
\end{equation}
where $\omega_\text{c}$ is the cyclotron frequency. We therefore find the ratio of the first harmonic between the gapped state and the normal state to be:
\begin{equation}
\frac{M_\text{g}}{M_\text{n}} = \frac{\sinh(X)}{X} \int_0^\infty \cos\left( \frac{X \mu}{\pi} \right) \partial_\mu \left( \frac{\mu}{\sqrt{\mu^2 + (\pi \delta / X)^2}} \tanh\left( \frac{\sqrt{\mu^2 + (\pi \delta / X)^2}}{2}\right) \right) \text{d}\mu.
\end{equation}
Gapped model simulations of the non-LK form of quantum oscillation amplitude at low temperatures are shown in the lower inset to Fig.~4a for various gap sizes (i.e. various sizes of $\delta$), compared with the LK growth in quantum oscillation amplitude at low temperatures for gapless models (i.e. $\delta=0$).

\subsubsection*{Model comparisons with experimental data}
Upper insets to Fig.~4a and Fig.~4b of the main text show the growth in quantum oscillation amplitude of the 700~T frequency in magnetic torque and 800~T frequency in electrical resistivity plotted against $X^2$, respectively, in the unconventional insulating phase of YbB$_{12}$. The LK exponential low temperature growth of the measured quantum oscillation amplitude observed for both electrical transport and torque magnetisation is in striking contrast to the non-LK finite temperature activation expected for gapped models of quantum oscillations (lower inset to Fig.~4a). \\[-6pt]

\noindent For the insulating regime of YbB$_{12}$ in which temperature dependent quantum oscillations are measured, the isotropic gap size at 40~T is given by $2\Delta\approx$~15~K~\cite{Sugiyama57.3946}, which yields $\delta \approx 12$ for $m^\ast/m_{\rm e}=7$ for the quantum oscillation frequencies shown in Fig.~4. Simulations with various values of $\delta$ are shown in the lower inset of Fig.~4a in the main text~\cite{Miyake186.115, Zhang116.046404, Knolle115.146401}. For the gapless case ($\delta = 0$), quantum oscillation amplitude simulations show an exponential LK growth at low temperature, while for the gapped case (finite $\delta$, shown for values up to $\delta = 10$, similar to YbB$_{12}$), quantum oscillation amplitude simulations show non-LK finite activation behaviour at low temperature. A comparison of measured quantum oscillation amplitude growth at low temperature with model simulations thus evidences neutral gapless excitations in the unconventional insulating phase of YbB$_{12}$.

\subsubsection*{Low temperature model expansion}

A further simplification may be yielded at low temperatures by using a low temperature expansion. We perform a series expansion of the term $\sinh(X)/X$ corresponding to the temperature damping term $R_\text{T}$ in the Lifshitz-Kosevich (LK) formula that describes the temperature dependence of quantum oscillations for particles obeying the Fermi-Dirac distribution~\cite{Shoenberg1984}. \\[-6pt]

\noindent For small $T$, a series expansion of the temperature dependence term yields:
\begin{equation}
    R_\text{T} \approx 1 - \frac{X^2}{6} + \mathit{O}\left(X^4\right).
\end{equation}
The quantum oscillation amplitude therefore linearly increases with decreasing $X^2$ approaching the zero $T$ limit. The low temperature growth in quantum oscillation amplitude is captured by the relative change of quantum oscillation amplitude at a finite temperature $A(T)$ with respect to the amplitude at the lowest measured temperature $A_0$, given by:
\begin{align}
    1 - \frac{A(T)}{A_0} &= \frac{A_0 - A(T)}{A_0} \nonumber \\
    &= \frac{X^2}{6}.
\end{align}
A plot of $(A_0 - A(T))/A_0$ against $X^2$ would therefore yield a straight line with a gradient equal to $1/6$ at low temperatures for low-energy excitations within the gap. In contrast, in the absence of low-energy excitations, gapped quantum oscillation models would yield a much reduced change in amplitude as a function of $X^2$ at low temperatures well below the gap temperature scale (Lower inset to Fig.~4a)~\cite{Miyake186.115, Zhang116.046404, Knolle115.146401}. A simplified comparison to distinguish between gapless and gapped forms of measured quantum oscillation amplitude is thus provided by this low temperature expansion.

\end{document}